\documentclass[12pt]{iopart}
\pdfoutput=1
\usepackage{graphicx}
\usepackage[latin1]{inputenc}
\usepackage{amsfonts}
\usepackage{amssymb}
\usepackage{amsthm}
\theoremstyle{definition} 
\theoremstyle{plain} 
\theoremstyle{remark} 
\graphicspath{{./images/}}
\begin{document}
\title[Dual Gardner problem]{The dual of the space of interactions in  neural network models}
\author{D.De Martino$^{1}$}
\address{$^1$ Center for life nanoscience, Istituto Italiano di Tecnologia, CLNS-IIT,  Viale Regina Elena 291, 00161, Rome, Italy}
\begin{abstract}
In this work the Gardner problem of inferring interactions and fields for an Ising neural network from given patterns under a local stability hypothesis  is addressed under a dual perspective. By means of duality arguments    
an integer linear system is defined whose solution space is the dual  of the Gardner space and whose solutions represent mutually unstable patterns. We propose and discuss Monte Carlo methods in order to find and remove unstable patterns and uniformly sample the space of interactions thereafter. We illustrate the problem on a set of real data and perform  ensemble calculation that shows how the emergence of phase dominated by unstable patterns can be triggered in a non-linear discontinuous way.  
\end{abstract}


\section{Introduction}
The use of concepts, methods and models from statistical mechanics, in particular of disordered systems and spin glasses, for the analysis of neural networks\cite{hopfield1982neural} and associative memories has a long standing tradition and has given many fruitful insights\cite{amit1992modeling}\cite{coolen2005theory}
\cite{agliariwalk}\cite{agliari2014hierarchical}\cite{barra2012glassy}. 
Recent years witnessed a renewed  surge of interest in modeling real biological neural networks with statistical mechanics models since it has been shown that experimental observations on the statistics of spiking patters can be reproduced within maximum-entropy Ising models\cite{schneidman2006weak}.
The computationally difficult inverse problem of inferring interactions and fields from patterns statistics has been addressed in several works(\cite{ricci2012bethe} and references therein), revealing many subtleties
concerning for instance the assessment of the extrapolated criticality of the inferred model\cite{mastromatteo2011criticality} \cite{marsili2013sampling}.
In general the problem of finding the right parameters of the model in order to store given patterns and structure them in an associative memory is an interesting problem of neural network learning.
In a different and simpler setting with respect to the aforementioned inverse problem, interactions and fields could be inferred under a local stability hypothesis of the patterns, that is the Gardner problem we will afford here\cite{gardner1988space}. 
Consider the patterns $\xi_{\mu i}=\pm 1$, where $\mu=1 \dots M$ is the pattern index and $i=1 \dots N$ is the neuron index.  
The pattern $\mu$ is stable if interactions $J_{ij}$(we consider a symmetric model $J_{ij}=J_{ji}$ without self interactions $J_{ii}=0$) and fields $h_i$ are such that the system of inequalities
\begin{equation}
\xi_{\mu i}(\sum_{j\neq i}J_{ij}\xi_{\mu j}+h_i)>0 \quad \forall i
\end{equation}
has solutions (the Hopfield choice $J_{ij}=\xi_{\mu i}\xi_{\mu j} \quad h_i =\xi_{\mu i}$ gives  a straightforward solution). All the patterns are mutually stable if the inequalities are verified $\forall \mu$ (in this case the Hopfield choice is not guaranteed to give a solution).
This is a linear system of inequalities for the $J_{ij}$ and $h_i$:  
\begin{equation}
\sum_{j\neq i}J_{ij}\xi_{\mu i}\xi_{\mu j}+h_i \xi_{\mu i}>0 \quad \forall \mu,i.
\end{equation}
The problem of finding interactions and fields and characterize statistically the space for given ensembles of patterns has been addressed in several works\cite{gardner1988optimal}\cite{krauth1987learning} \cite{mezard1989space}  . 

In this work the Gardner problem   is afforded with a dual perspective.
From duality arguments we define an integer linear  system whose solutions give a certificate of the infeasibility of the Gardner problem and represent sets of mutually unstable patterns. In the following we discuss Montecarlo methods in order to find such sets and sample the space of interactions thereafter, we analyze with them a simple instance of biological data and perform ensemble calculations. We finally conclude summarizing the results and drawing some future perspectives on this problem.

\section{Results}

\subsection{The dual Gardner problem} 
We have seen that a local stability hypothesis for the patterns imply that the interactions and fields of an Ising neural network should verify a system of linear inequalities. 
The existence of solutions for a system of linear inequalities is ruled by duality theorems  that state that
the system has solutions if and only if a suitably defined dual system has no solutions\cite{schrijver1998theory}.
The space of interactions and fields, system (2), is defined by $NM$ inequalities for $N(N+1)/2$ variables. 
According to the Gordan theorem the dual of system (2) is
\begin{eqnarray}
\sum_\mu k_{\mu i} \xi_{\mu i} \xi_{\mu j}=0 \quad \forall i<j \\
\sum_\mu k_{\mu i} \xi_{\mu i}=0 \quad \forall i  \nonumber \\
k_{\mu i}\geq 0 \nonumber
\end{eqnarray}
that is a system of $N(N+1)/2$ linear equations for the $NM$ non negative variables $k_{\mu i}$. 
In the appendix we report a self consistent demonstration of the theorem based on the Fourier-Motzkin-Chernikova elimination method.
It should be noticed that the matrices defining the constraints are integers and thus any solution can be represented in terms of integer solutions, and in essence we can consider system (3) as an integer linear system. Further we note that the equations do not mix variables $k_{\mu i}$ with different $i$ and $j$.  In the following section we will discuss numerical methods to find solutions  for both systems, the space of interactions and fields and its dual.

\subsection{Resolution methods}
A system of linear inequalities like system (2) can be solved by relaxational algorithms. Starting from a prior $(J_{ij}^0, h_i^0)$, we calculate the least unsatisfied constraint from (2) and update the parameters along the orthogonal direction, defining the series
\begin{eqnarray}
(\mu_t,i_t)=\textrm{argmin}_{\mu,i} \sum_{j\neq i}J_{ij}^t\xi_{\mu i}\xi_{\mu j}+h_i^t \xi_{\mu i} \\
J_{i_{t}j}^{t+1}=J_{i_{t}j}^t+\gamma \xi_{\mu_{t} i_{t}} \xi_{\mu_{t} j} \quad h_{i_{t}}^{t+1}=h_{i_{t}}^t+\gamma \xi_{\mu_{t} i_{t}} 
\end{eqnarray}
that will converge in polynomial time to a solution, provided its existence. The length $\gamma$ of the step can be constant (MinOver\cite{krauth1987learning}) or proportional to the amount by which the constraint is violated (Motzkin algorithm\cite{motzkin1954relaxation}).
In absence of solutions the learning through relaxation does not  converge, and according to the Gordan theorem there should be solutions to the dual system (3).
On the other hand, the solutions of the dual system (3) can be seen as the ground state with zero energy of the Hamiltonian over integer variables obtained summing over the constraints square:
\begin{equation}
H = \frac{1}{N}\sum_{i<j} (\sum_\mu k_{\mu i} \xi_{\mu i} \xi_{\mu j})^2 +\frac{1}{N}\sum_i (\sum_\mu k_{\mu i} \xi_{\mu i})^2
\end{equation}
In this way it is possible to use computational methods from statistical mechanics to find  solutions.
For instance, upon introducing a fictitious inverse temperature $\beta$ we can sample equilibrium configurations from the Boltzmann distribution $P\propto e^{-\beta H}$ with a Metropolis-Hastings  Montecarlo Markov chain and decrease the temperature (simulated annealing) till convergence to a ground state configuration (eventually the trivial solution). In order to reduce the dimension of the space we can couple the methods\cite{de2013counting}, i.e. we can search for solutions of system (3) only among the variables that correspond to most unsatisfied constraints for a previous search of solutions to system (2). In this way we will have either a solution to the Gardner problem or a set of mutually unstable patterns that proves that the Gardner problem is unfeasible for such patterns. In the latter case we could reject and remove one of the pattern in the set and start again. The choice of the pattern to remove in the set should depend on the specific context of the problem afforded. 
If, on the other hand, the algorithm converge to a solution of the Gardner problem, a complete characterization of the solution space of system (2), that is geometrically a polyhedral cone,  would require the calculation of a so called Hilbert basis\cite{henk1996hilbert}, i.e. the complete set of solutions that cannot be represented as convex sums of other solutions. This is a difficult computational problem and  an alternative is to characterize the space spanned by solutions of system (2) with statistical methods. This would require to fix a scale for the space and a simple choice consists in bounding interactions and fields in way that depends on the specific context of the problem afforded, for instance $J_{ij}\in[-C,C]$ $h_i\in[-C,C]$.  
In this way the space of interactions form a convex polytope. An uniform sampling of convex space in high dimensions can be performed in feasible times by means of Monte Carlo Markov chains, the Hit and Run method being the faster so far\cite{HRb} \cite{lovasz1999hit}. 
Given a $D$-dimensional convex set $P$, from which one wants to sample from, and a point inside $x_t \in P$, the Hit and Run algorithm is defined as follows:
\begin{enumerate}
  \item Choose a uniformly distributed direction $\theta_t$, that is, a point extracted from the uniform distribution on the $D$-dimensional unit sphere. This can be done with the Marsaglia method, i.e. by generating $D$ independent gaussian random variables $\theta_t^i$ with zero mean and unit variance, and then normalizing the vector to unit length;
  \item Extract $\lambda^\star$ uniformly from the interval $[\lambda_{\min},\lambda_{\max}]$, where $\lambda_{\min}$ ($\lambda_{\max}$) is the minimum (maximum) value of $\lambda$ such that $x_{t} + \lambda \theta_t \in P$;
  \item Compute the point $x_{t+1} = x_{t} + \lambda^\star \theta_t$, increment $t$ by one and start again.
\end{enumerate}
The mixing time $\tau$ of the hit and run scales like\cite{lovasz1999hit}
\begin{equation}
\tau \simeq \mathcal O(D^2 R^2/r^2)
\end{equation} 
where $D$ is the dimension of the polytope, $R,r$ are the radii  of respectively the  minimum inscribing  and the maximum inscribed balls.
The factor $R/r$ can be large and leading to ill-conditioning but it can be reduced to 
$\sqrt D$ for centrally s{y}mmetric polytopes and to  $D$ in general, by an affine transformation defined by 
the so-called {Loewner--John} Ellipsoid\cite{ball1992ellipsoids}, i.e. the ellipsoid of maximal volume contained in the polytope.
Unfortunately this ellipsoid cannot be found in polynomial time, but it has been shown by {L.~Lovazs} that a weaker form of the { Loewner--John} ellipsoid, with a factor of $D^{3/2}$, can be found in polynomial time\cite{lovasz1987algorithmic}.

\subsection{Analysis of real data: an illustrative example}
In this section we will analyze a subset of  experimental data on the neurons spiking patterns in a salamander retina, in particular we refer to the snapshot of $T=2s$ reported in Fig 1 of\cite{tkavcik2014searching}. There are $M=31$ distinct patterns for $N=15$ effectively spiking neurons that we report in the appendix. The relaxational algorithm applied to the space of synaptic strengths for this network does not converge indicating the presence of conflicting patterns that have been identified with a Montecarlo simulated annealing. For sake of simplicity, we have chosen to remove one of the pattern picked uniformly at random in the retrieved set. 
We report in the appendix the set of conflicting patterns retrieved in one run of the algorithm and the patterns we have removed. Once we have removed a sufficient number of conflicting patterns such that the system (2) defining the space of synaptic strength is feasible we can sample interactions and fields.  
\begin{figure}[h!!!!]
\begin{center}
\includegraphics*[width=.65\textwidth]{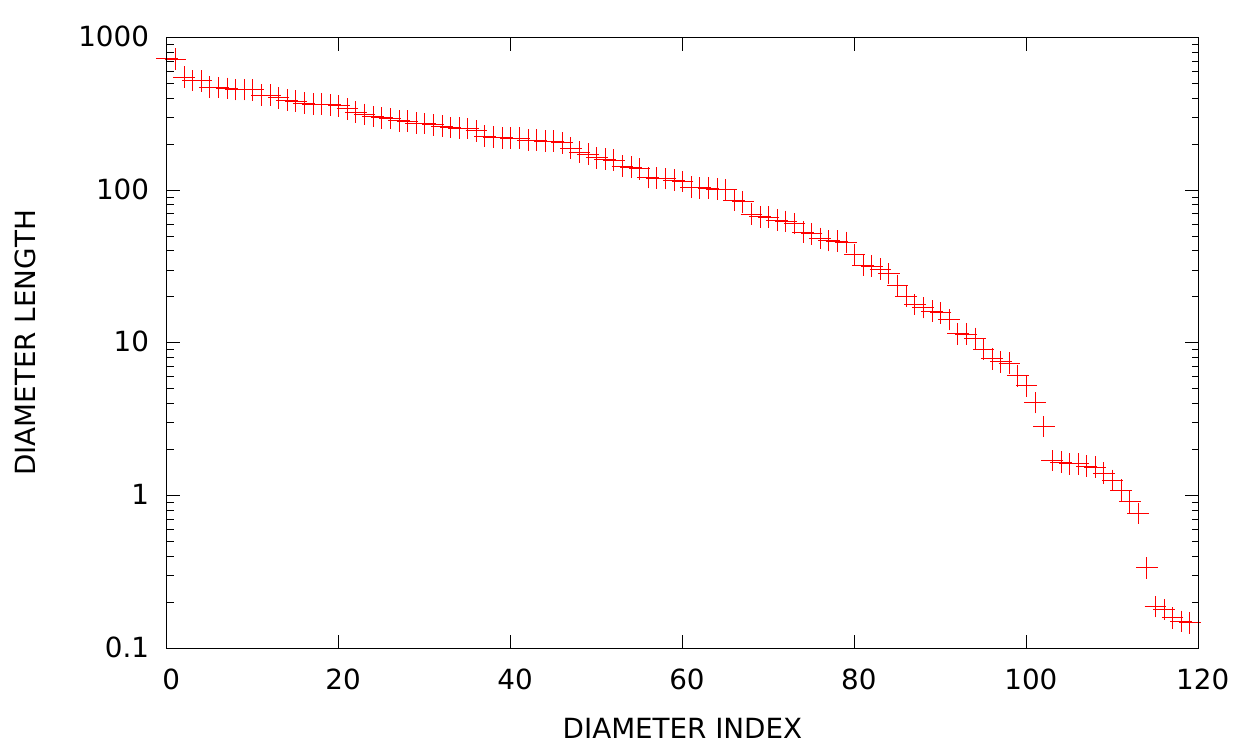}
\caption{diameters length of the ellipsoid that rounds the polytope representing the space of  neural network models storing a given set of stable patterns (see appendix).}
\end{center}
\end{figure}
We have chosen to bound $J_{ij}\in[-1000,1000]$ $h_i\in[-1000,1000]$ and we have sampled uniformly the space by means of the Hit-and-run Markov chain after ellipsoidal rounding preprocessing\cite{uniformell}.
In Fig 1  we show the length of the diameters of the ellipsoid that rounds the polytope  (the dimension is $D=N(N+1)/2=120$) for decreasing order, where we can appreciate the heterogeneous structure of the space (the ratio between the largest and smallest diameter is of order $10^4$).
\begin{figure}[h!!!!]
\begin{center}
\includegraphics*[width=1\textwidth]{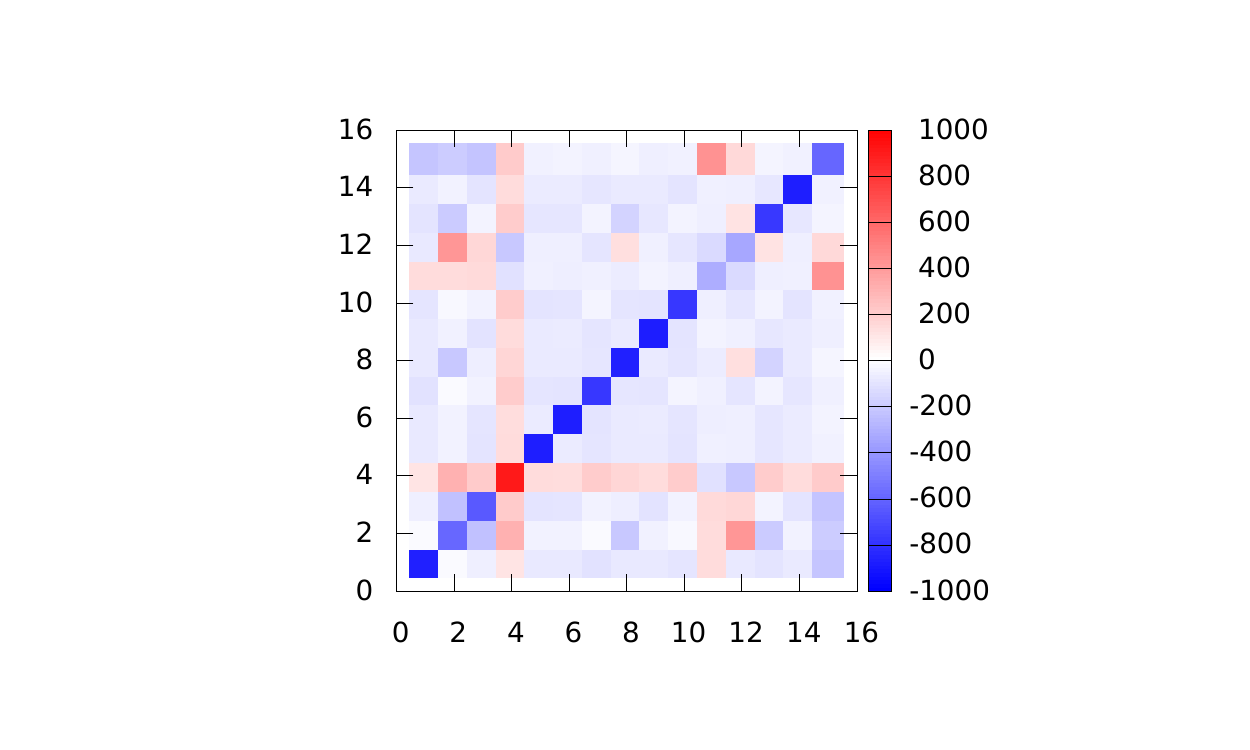}
\caption{Average interaction strengths and fields retrieved from uniform sampling of the space of neural network models corresponding to the stable set of patterns reported in the appendix.}
\end{center}
\end{figure}
In Fig 2  we show the averages of the interaction strengths (off-diagonal) and fields (diagonal) thus retrieved.   Fields are on average negative for all neurons apart from the fourth that has positive (ferromagnetic) interactions with the other neurons (whose interactions are on average weakly anti ferromagnetic)

\subsection{Ensemble calculations in a simple setting}
In order to get a picture of the dependence on $N$ and $M$ 
of the volume of the solution space of (3), the problem is that, at odds with linear algebra, for integer linear systems there are no straightforward results relating it for instance to the kernel of the constraints matrix.   
In order to get such insights we will perform here  a statistical analysis of the solution space of (3) by means of ensemble calculations in a simple setting.  
We will consider for simplicity an  integer linear system for the variables $k_\mu=0,1$:
\begin{equation}
\sum_\mu k_\mu \xi_{\mu i}=0  \qquad \forall i
\end{equation} 
with $\mu=1\dots M$, $i=1\dots N$. We want to enumerate solutions for random systems with $\xi_{\mu i}=\pm 1$ with equal probability  (a maximum entropy case). They are the ground states with zero energy of the system with hamiltonian
\begin{equation}
H = \frac{1}{N}\sum_i (\sum_\mu k_\mu \xi_{\mu i})^2.
\end{equation}
We would like to evaluate  $\overline{\log Z}$, where $Z=\sum_{k_\mu=0,1} e^{-\beta H}$ and the bar stands for the average over disorder.
We will  calculate $\overline{Z^n}$, and perform the limit
\begin{equation}
\overline{\log Z} =\lim_{n \to 0} \frac{1}{n}\log \overline{Z^n}. 
\end{equation}
Then we will send  $\beta \to \infty$, and $N,M\to \infty$ with $\alpha =\frac{M}{N}$ finite. 
We have  ($a=1 \dots n$ is the replica index)
\begin{eqnarray}
Z^n =\sum_{k_{\mu a}} \prod_{j a} e^{-\beta/N (\sum_\mu k_{\mu a} \xi_{\mu j})^2} = \nonumber \\ 
= \sum_{k_{\mu a}} \prod_{\mu \nu} e^{-\beta \sum_a k_{\mu a} k_{\nu a} \left( \frac{1}{N}\sum_j \xi_{\mu j} \xi_{\nu j}\right)}
\end{eqnarray}
where we have simply developed the square and exchanged index
in the sums and products. Now we observe that for $N$ large
\begin{equation}
  \frac{1}{N}\sum_j \xi_{\mu j} \xi_{\nu j} = \left\{ \begin{array}{rlc}
  1 & \mbox{if} & \mu=\nu\;,\\
  \to x_{\mu \nu} & \mbox{if} & \mu\neq \nu\;.
\end{array} \right.
\end{equation}
where $x_{\mu \nu}$ are independent gaussian random variables centered in $0$ with std $\frac{1}{\sqrt{N}}$. Then we have 
\begin{eqnarray}
\overline{Z^n} =\sum_{k_{\mu a}} \int \prod_{\mu < \nu} \frac{dx_{\mu \nu}}{\sqrt{2\pi/N}} e^{-2\beta \left(\sum_a k_{\mu a} k_{\nu a}\right) x_{\mu \nu}-N\frac{x_{\mu  \nu}^2}{2}} \prod_{\mu} e^{-\beta \sum_a k_{\mu a} k_{\mu a}} = \nonumber \\ 
= \sum_{k_{\mu a}} \prod_{\mu \neq \nu} e^{\beta^2/N (\sum_a k_{\mu a} k_{\nu a})^2} \prod_{\mu} e^{-\beta \sum_a k_{\mu a} k_{\mu a}}=\nonumber \\
=\sum_{k_{\mu a}} \prod_{a,b} e^{\beta^2/N \left(  (\sum_\mu k_{\mu a} k_{\mu b})^2 - \sum_\mu k_{\mu a} k_{\mu b} \right)}\prod_a e^{-\beta \sum_\mu k_{\mu a} k_{\mu a}}
\end{eqnarray}
where in the last line we have used $k_{\mu a}^2=k_{\mu a}$, however the term for the correction $\mu \neq \nu $ is not extensive and  we will drop it in the following.
We use known property of Gaussian integrals:
\begin{equation}
e^{\frac{1}{N} (\beta \sum_\mu k_{\mu a} k_{\mu b})^2}=\int dQ_{ab} \frac{e^{-N \frac{Q_{ab}^2}{4}+\beta Q_{ab} \sum_\mu k_{\mu a} k_{\mu b}}}{\sqrt{4\pi/N}} 
\end{equation}
we have then
\begin{eqnarray}
\overline{Z^n} = \int dQ \prod_{a,b}\frac{ e^{-N \frac{Q_{ab}^2}{4}}}{\sqrt{4\pi/N}}   \big( \sum_{k_a} \prod_{a,b} e^{\beta k_a Q_{ab} k_b} \prod_a e^{-\beta k_a}  \big)^M = \nonumber \\
=\int dQ e^{M F(Q)} \nonumber \\
F(Q) = -\frac{1}{4\alpha}\sum_{a,b} Q_{ab}^{2}+\log Z_r(Q) \nonumber \\
Z_r = \sum_{k_a} e^{-\beta H_r} \qquad H_r =-\sum_{ab} k_a Q_{ab} k_b+\sum_a k_a 
\end{eqnarray}
We see that saddle point equations lead to the following interpretation of the $Q_{ab}$:
\begin{equation}
\frac{\partial F}{\partial Q_{ab}}= 0 \quad \Rightarrow \quad  Q_{ab}^{SP}=2\alpha\beta \langle k_a k_b \rangle
\end{equation}  
where the brackets stand for averages among interacting replicas with  hamiltonian $H_r$.
A natural ansatz is to assume replica-symmetric matrices
\begin{equation}
  Q_{ab} = \left\{ \begin{array}{rlc}
  q_0 & \mbox{if} & a=b\;,\\
  q_1  & \mbox{if} & a\neq b\;.
\end{array} \right.
\end{equation}
and we have the expression
\begin{eqnarray}
F(q_0,q_1)  =-\frac{n}{4\alpha}(q_0^2+(n-1)q_1^2)+\log Z_r \nonumber \\
Z_r = \frac{1}{\sqrt{2\pi}} \int d\lambda e^{-\frac{\lambda^2}{2}}\big( 1+e^{\sqrt{2\beta q_1}\lambda-\beta(1+q_1-q_0)}\big)^n
\end{eqnarray}
where we have been using again known properties of gaussian integrals. 
The free energy $f$ in the limit $n \to 0$
\begin{eqnarray}
-\beta f=\lim_{n \to 0} F/n = \nonumber \\
=-\frac{1}{4\alpha}(q_0^2-q_1^2)+\frac{1}{\sqrt{2\pi}} \int d\lambda e^{-\frac{\lambda^2}{2}}\log(1+e^{ \sqrt{2\beta q_1}\lambda-\beta(1+q_1-q_0) })
\end{eqnarray}
If we analytically continue the SP equations
for $n \to 0$ we obtain:
\begin{eqnarray}
q_0 =2\alpha\beta \frac{1}{\sqrt{2\pi}} \int d\lambda e^{-\frac{\lambda^2}{2}} \frac{1}{1+e^{ -\sqrt{2\beta q_1}\lambda+\beta(1+q_1-q_0) }} \nonumber \\
q_1=q_0-2\alpha \sqrt{\frac{\beta}{2 q_1}}\frac{1}{\sqrt{2\pi}} \int d\lambda e^{-\frac{\lambda^2}{2}} \frac{\lambda}{1+e^{ -\sqrt{2\beta q_1}\lambda+\beta(1+q_1-q_0) }} 
\end{eqnarray}
In the limit $\beta \to \infty$ we look at the solution 
\begin{eqnarray}
\lim_{\beta \to \infty} q_0/\beta = \lim_{\beta \to \infty} q_1/\beta = x \nonumber \\
x = 2\alpha \frac{1}{\sqrt{2\pi}} \int_{\frac{1}{\sqrt{2x}}}^{\infty} d\lambda e^{-\frac{\lambda^2}{2}}  \\
f=-\frac{1}{\sqrt{2\pi}} \int_{\frac{1}{\sqrt{2x}}}^{\infty} d\lambda e^{-\frac{\lambda^2}{2}}(\sqrt{2x}\lambda-1)
\end{eqnarray}
We remind that the original problem is for $NM$ variables and $N(N-1)/2$ equations and thus we should consider $\alpha/2$ as the correct control parameter, further upon looking at the interpretation of the saddle point equations, we will consider
$r = 2\alpha x$ as an order parameter in $(0,1)$ 
Apart from  the trivial solution $r=0$  we depict in the following figure the curves  $r(\alpha)$ and $f(\alpha)$.
\begin{figure}[h!]
\begin{center}
\includegraphics*[width=.45\textwidth]{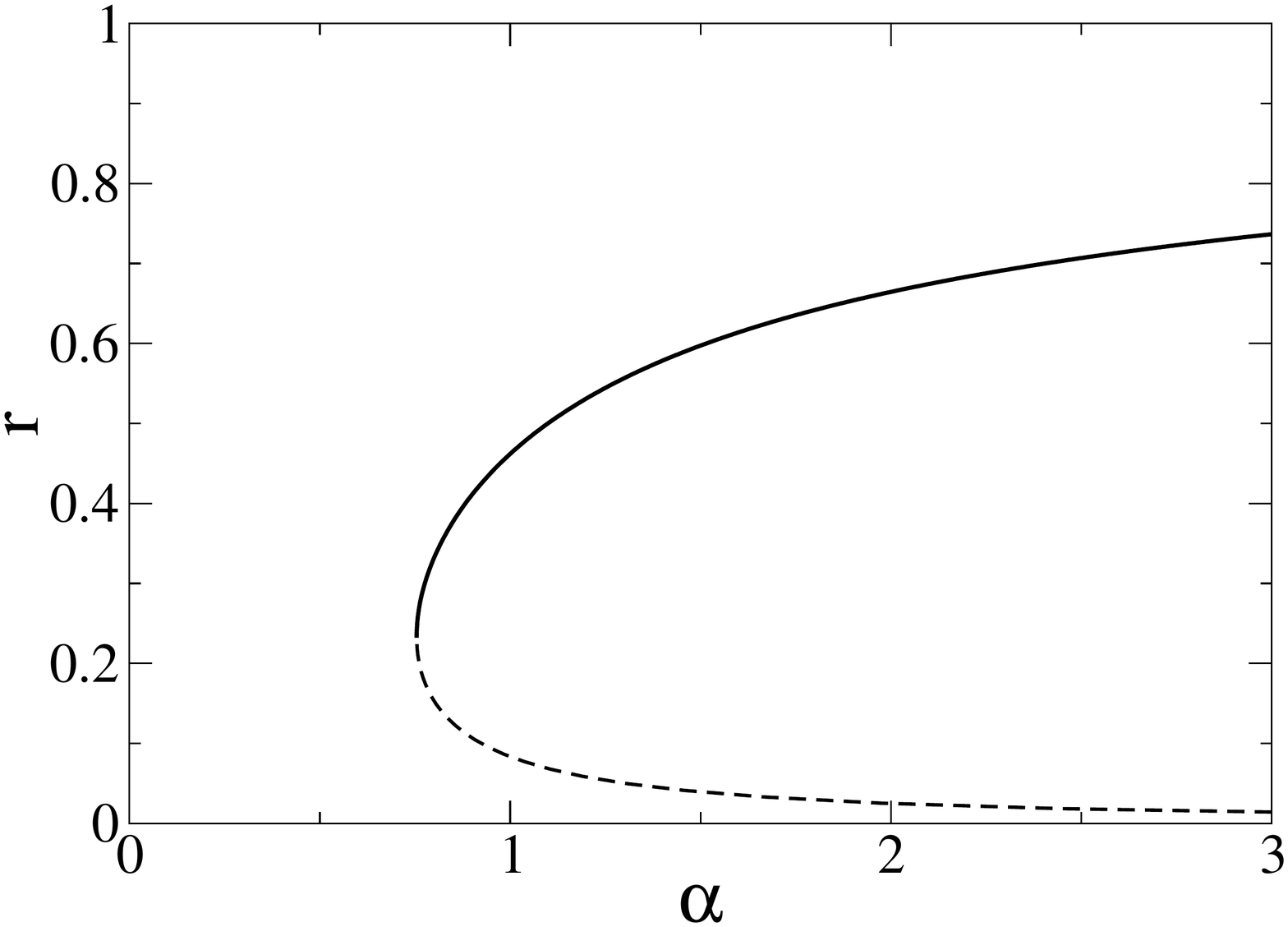}
\includegraphics*[width=.45\textwidth]{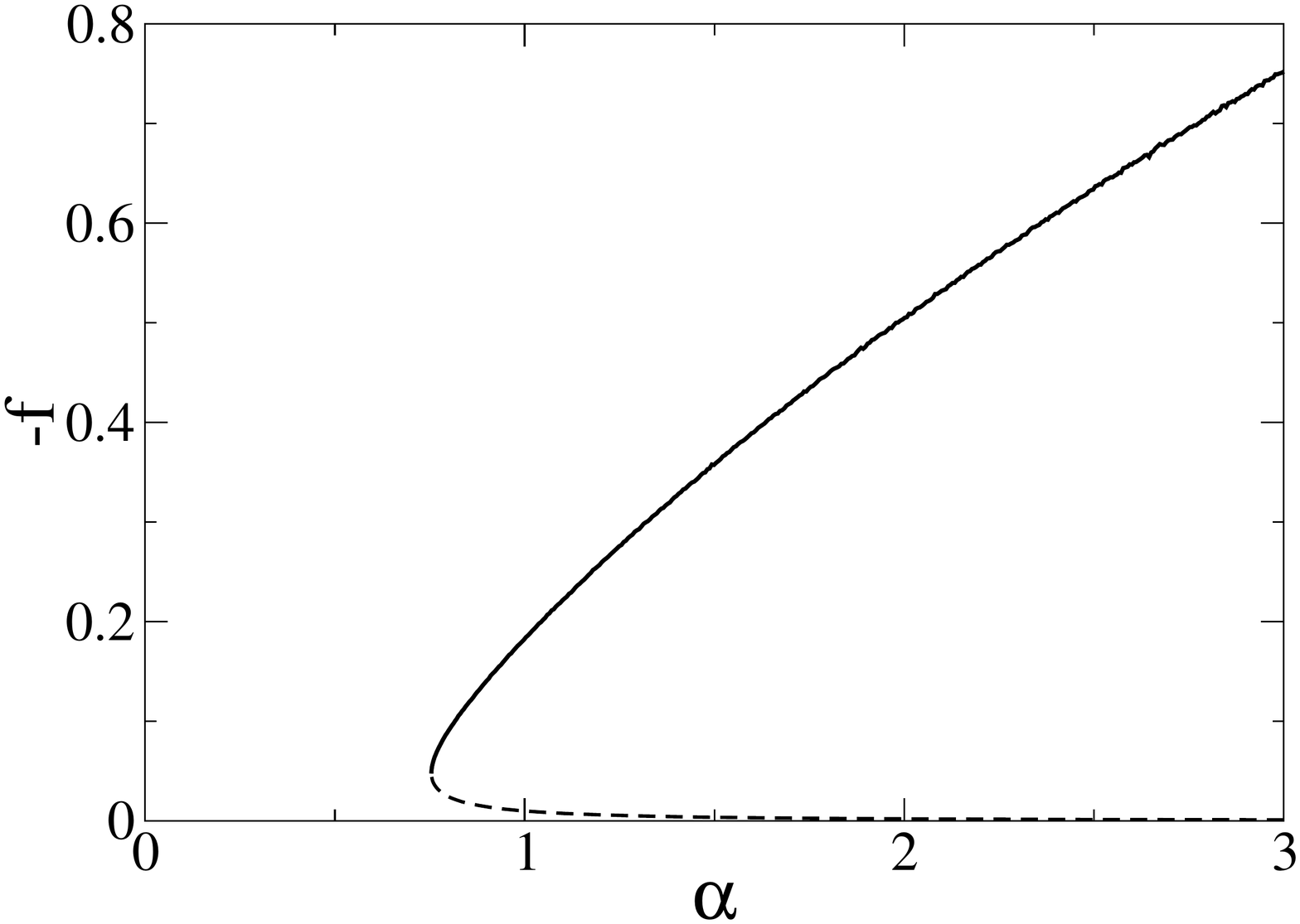}
\caption{Order parameter $r$ and (minus) free energy $f$ as a function of $\alpha=M/N$ from replica symmetric ensemble calculations.}
\end{center}
\end{figure}

\section{Conclusions}
In this work we have defined the dual of the space of interactions of neural network models under a stability hypothesis for the patterns. By means of the Gordan theorem we derived an integer linear system whose solutions represent mutually unstable patterns whose feasibility implies the infeasibility of the Gardner problem.  We have discussed Monte Carlo methods for detecting and removing such instabilities and finally perform an uniform sampling of the space of interactions and fields. We have applied them for illustrative purposes to the analysis of a simple data set of spiking patterns from biological neurons. We have performed replica-symmetric ensemble calculations in a simple setting showing that the emergence of a phase dominated by unstable patterns can be triggered in a discontinuous way.  
Apart from an analysis of more complex ensembles it would be important to perform  stability analysis, finite temperature calculations and for for general integer variables $k_i=0,1,\dots, m$. Recent analysis seems to show however that for sparse random integer linear system the behavior for $m>1$ is qualitatively  similar to that for $m=1$ and that the replica symmetric solution is stable\cite{simona}.  
Regarding applications, for instance in analyzing real data of spiking neurons, we stress that our choice to remove a pattern uniformly at random among the ones retrieved in unstable sets is motivated by simplicity and for illustrative purposes.  In order to implement different choices it would be worth to have an overall picture of the space of mutually unstable patterns for single instances, beyond our Montecarlo method that retrieves just one solution. Probably this problem could gain fruitful insights from message passing and cavity methods, that have been recently applied  to integer linear systems but in a different context\cite{font2012weighted}
\cite{braunstein2008estimating} \cite{de2014identifying}.

\section*{Appendix}
\subsection*{Demonstration of the Gordan theorem}
Consider the system of homogeneous linear inequalities for the variables $x_\mu$  defined by the matrix $A_{i \mu}$
\begin{equation}
\label{sys1}
\sum_\mu A_{i \mu} x_\mu > 0 \qquad \forall i.
\end{equation}.
We associate  it to the following dual system for the non-negative variables $y_i$   
\begin{eqnarray}
\label{sys2}
\sum_i A_{i \mu} y_i = 0 \qquad \forall \mu  \nonumber \\
y_i\geq 0, \quad {\bf y}\neq {\bf 0}
\end{eqnarray}
We want to demonstrate the Gordan theorem:\\
One and only one of the systems (\ref{sys1}) and (\ref{sys2}) have solution.\\
We first demonstrate that\\
(\ref{sys2}) has solution $\Rightarrow$ (\ref{sys1}) has no solution.\\
Suppose there is a solution ${\bf y^*}$ of (\ref{sys2}). Take any vector ${\bf x}$, multiply it component by component
with the equations of the system (\ref{sys2}) and sum over them. Exchanging the indeces $\mu$ and $i$ 
in the sums and given the positivity of ${\bf y^*}$, 
it is straightforward to conclude that no vector ${ \bf x}$ can satisfy the system (\ref{sys1}).\\ 
Now we demonstrate that\\
(\ref{sys1}) has no solution $\Rightarrow$ (\ref{sys2}) has solution.\\
The demonstration is given by induction in the number $M$ of unknowns $x_\mu$. 
The statement is true for $M=1$. Infact, for one unknown, the system (\ref{sys1}) is inconsistent  
if and only if there is at least one couple  $i$, $j$ such that $A_{i 1} A_{j 1} = -1/c < 0$, 
and in this case $y_i=1$, $y_j=c$ and $y_l=0 \quad \forall l\neq i,j$ is a solution of (\ref{sys2}).    
Let's consider a system of the type (\ref{sys1}) with $M$ unknowns and suppose it is inconsistent.
We will prove that the dual system has solutions supposing the theorem true for systems with $M-1$ unknowns.     
We have $\forall i \quad \sum_{\mu=1}^{M-1} A_{i \mu} x_\mu > -A_{i M} x_M$, 
 if $A_{i M} \neq 0$, we can define $\tilde{A}_{i \mu} = - A_{i \mu}/A_{i M}$, and we have:
\begin{eqnarray}
\sum_{\mu}^{M-1}\tilde{A}_{i \mu} x_\mu = P_i > x_M  \qquad  \forall i: \quad A_{i M}<0 \nonumber \\
\sum_{\mu}^{M-1}\tilde{A}_{j \mu} x_\mu = Q_j < x_M  \qquad  \forall j: \quad A_{j M}>0 \nonumber \\
\sum_{\mu}^{M-1}A_{l \mu} x_\mu = R_l > 0   \qquad \forall l: \quad A_{l M}=0.
\end{eqnarray}  
Writing the system in this form, we can pass to the following system in $M-1$ unknows:
\begin{eqnarray}
\label{sysprof}
P_i> Q_j \qquad \forall i,j: \quad A_{i M}<0 \quad A_{j M}>0 \nonumber \\
R_l > 0   \qquad \forall l: \quad A_{l M}=0.
\end{eqnarray}
Now, this system is also inconsistent. \\ Suppose infact there is a solution ${\bf x^*} = ( x_\mu^* )$, $\mu=1 \dots M-1$.\\ 
We could add to it any $x_M^*$ such that   $max_j Q_j({\bf x^*}) < x_M^* <min_i P_i({\bf x^*})$ 
and we will have a solution for the original system as well, against the hypothesis.     
By induction, the theorem is true for systems with $M-1$ unknowns. 
Then, referring to (\ref{sysprof}), there are $\tilde{y}_{ij}\geq 0$, $y_l \geq 0$ with at least one positive,  
such that $\sum_{ij} \tilde{y}_{ij} (\tilde{A}_{i \mu} - \tilde{A}_{j \mu}) + \sum_l y_l A_{l \mu} = 0 \quad \forall \mu$.\\ From this we have finally a solution for the system (\ref{sys2}):
\begin{eqnarray}
y_i = - \sum_j \tilde{y}_{ij}/A_{i M}  \qquad  \forall i: \quad A_{i M}<0 \nonumber \\
y_j = \sum_i \tilde{y}_{ij}/A_{j M} \qquad   \forall i: \quad A_{j M}>0  \nonumber \\
y_l  \qquad    \forall l: \quad A_{l M}=0,
\end{eqnarray}
and the theorem is proven.
Regarding the application to the Gardner problem it is useful to remind that, essentially if we group index $r=(\mu,i)$, $s=(i,j)$ the matrix is
\begin{equation}
  A_{(\mu,i)(i,j)} = \left\{ \begin{array}{rlc}
  \xi_{\mu i} (0)& \mbox{with (without) fields if} & i=j \;,\\
  \xi_{\mu i}\xi_{\mu j} & \mbox{if} & i\neq j \;.
\end{array} \right.
\end{equation} 
\subsection*{Supplementary tables}
We report here the patterns analyzed in sec 2.3 taken from  \cite{tkavcik2014searching}, i.e. the $0-1$  matrix $n_{i \mu}$ where $n_{i \mu}=1$ ($0$) means that  the neuron $i$ is firing (not) in the pattern $\mu$.
\begin{equation}
\begin{array}{c c} &
\begin{array}{c} \textrm{NEURONS} \\
\end{array}
\\
\begin{array}{c c c c c c c c}
P \\
A \\
T \\
T \\
E \\
R \\
N \\
S \\
\end{array}
&
\left[
\begin{array}{c c c c c c c c c c c c c c c}
0 & 0 & 1 & 0 & 0 & 0 & 0 & 0 & 0 & 0 & 0 & 0 & 0 & 0 & 0 \\ 
1 & 0 & 0 & 0 & 0 & 0 & 0 & 0 & 0 & 0 & 0 & 0 & 0 & 0 & 0 \\ 
0 & 0 & 1 & 0 & 0 & 0 & 0 & 0 & 0 & 0 & 1 & 0 & 0 & 0 & 0 \\ 
0 & 0 & 0 & 0 & 0 & 0 & 0 & 0 & 0 & 1 & 0 & 0 & 0 & 0 & 0 \\ 
1 & 0 & 0 & 0 & 0 & 0 & 0 & 0 & 0 & 0 & 1 & 0 & 0 & 0 & 0 \\ 
0 & 0 & 0 & 0 & 0 & 0 & 0 & 0 & 0 & 0 & 0 & 1 & 0 & 0 & 0 \\ 
0 & 0 & 0 & 0 & 0 & 0 & 0 & 0 & 0 & 0 & 0 & 0 & 1 & 0 & 0 \\ 
0 & 0 & 1 & 0 & 0 & 0 & 0 & 0 & 1 & 0 & 0 & 0 & 0 & 0 & 0 \\ 
0 & 0 & 0 & 0 & 0 & 0 & 0 & 1 & 0 & 0 & 0 & 0 & 1 & 0 & 0 \\ 
0 & 0 & 0 & 0 & 0 & 0 & 0 & 1 & 0 & 0 & 0 & 0 & 0 & 0 & 0 \\ 
0 & 0 & 0 & 0 & 0 & 0 & 1 & 0 & 0 & 0 & 0 & 0 & 0 & 0 & 0 \\ 
0 & 0 & 0 & 0 & 0 & 1 & 0 & 0 & 0 & 0 & 0 & 0 & 0 & 0 & 0 \\ 
0 & 0 & 1 & 0 & 0 & 0 & 0 & 1 & 0 & 0 & 0 & 1 & 0 & 0 & 0 \\ 
0 & 0 & 1 & 0 & 0 & 0 & 1 & 1 & 0 & 0 & 0 & 0 & 0 & 0 & 0 \\ 
0 & 1 & 0 & 0 & 0 & 0 & 0 & 0 & 0 & 0 & 0 & 1 & 0 & 0 & 0 \\ 
0 & 0 & 0 & 0 & 0 & 0 & 0 & 0 & 1 & 0 & 0 & 0 & 0 & 0 & 0 \\ 
0 & 0 & 0 & 0 & 0 & 0 & 0 & 0 & 0 & 0 & 0 & 0 & 0 & 1 & 0 \\ 
0 & 0 & 0 & 0 & 1 & 0 & 0 & 0 & 0 & 0 & 0 & 0 & 0 & 0 & 0 \\ 
0 & 0 & 0 & 1 & 0 & 0 & 0 & 0 & 0 & 0 & 0 & 0 & 0 & 0 & 0 \\ 
1 & 0 & 1 & 0 & 0 & 0 & 0 & 0 & 0 & 0 & 1 & 0 & 0 & 0 & 0 \\ 
0 & 0 & 1 & 0 & 0 & 0 & 0 & 0 & 0 & 0 & 0 & 0 & 0 & 0 & 1 \\ 
0 & 0 & 0 & 0 & 0 & 0 & 0 & 0 & 0 & 0 & 1 & 0 & 0 & 0 & 1 \\ 
0 & 0 & 1 & 0 & 0 & 0 & 0 & 0 & 0 & 0 & 1 & 0 & 0 & 0 & 0 \\ 
0 & 0 & 0 & 0 & 0 & 0 & 0 & 0 & 0 & 0 & 1 & 0 & 0 & 0 & 0 \\ 
0 & 0 & 1 & 0 & 0 & 0 & 0 & 0 & 0 & 0 & 0 & 1 & 1 & 0 & 0 \\ 
0 & 0 & 0 & 0 & 0 & 0 & 0 & 0 & 1 & 0 & 0 & 0 & 1 & 0 & 0 \\ 
0 & 0 & 1 & 1 & 0 & 0 & 1 & 0 & 0 & 1 & 0 & 0 & 1 & 0 & 0 \\ 
0 & 0 & 1 & 0 & 0 & 0 & 0 & 0 & 0 & 0 & 0 & 0 & 1 & 0 & 0 \\ 
0 & 0 & 0 & 1 & 0 & 0 & 0 & 0 & 0 & 0 & 0 & 0 & 1 & 0 & 0 \\ 
0 & 0 & 0 & 0 & 0 & 0 & 1 & 0 & 0 & 0 & 0 & 0 & 0 & 0 & 0 \\ 
0 & 0 & 0 & 0 & 0 & 0 & 0 & 0 & 0 & 0 & 0 & 0 & 0 & 0 & 0 
\end{array}
\right]
\end{array}
\end{equation}
We have been using the usual transformation $\xi_{i \mu}=2n_{i \mu}-1$. We report in the following table 1 the unstable patterns solutions of system (3) retrieved and removed in one run of the algorithm described in sec 2.2.  The information given in the table should be interpreted in the following way: if patterns $29$ and $6$ are conflicting for neuron $4$, this means that system (3) has a solution $k_{4,29}=k_{4,6}=1$ and $k_{i \mu}=0$ for the other  components.      
    
\begin{table}[h!!!]
\begin{center}
\begin{tabular}{  | c  | c  |  c | }  
\hline 
Conflicting patterns & Neuron & Pattern removed  \\ 
\hline 
29,6  &  4  & 29 \\
30,31  &  7  & 31 \\
2,5  &  11  & 5 \\
9,10  &  13  & 9 \\
1,10,11,14  &  7  & 14 \\
16,26  &  13  & 26 \\
20,23  &  1  & 23 \\
7,28  &  3  & 28 \\
8,16  &  3  & 8 \\
1,6,7,25  &  12  & 6 \\
4,19  &  4  & 19 \\
1,2,20,24  &  11  & 24 \\
21,22  & 11  & 21 \\
3,20  &  1  & 3 \\
\hline
\end{tabular}
\caption{First Column: Set of conflicting patterns. Second Column: Reference neuron. Third column: Removed pattern.
Obtained by coupling Relaxation and Montecarlo simulated annealing applied to the data represented in matrix (20).}
\end{center}
\end{table}

\section*{Acknowledgments}
I would like to warmly thank Luca Leuzzi for fruitful discussions on the replica method.    
\section*{References}    

\bibliographystyle{unsrt}
\bibliography{refDual}

\end{document}